\documentclass[twocolumn,aps,prb]{revtex4}
\usepackage{graphics,graphicx}
\usepackage{dcolumn}
\usepackage{bm}

\newcommand{\Ref}[1]{Ref.~\onlinecite{#1}}

\def\eb{\begin{equation}}   
\def\ee{\end{equation}}     
\def\ea#1{\begin{eqnarray} #1 \end{eqnarray}}   

\def\shro{Schr\"odinger}

\def\ra{\rightarrow}

\def\im{\text{Im}}
\def\re{\text{Re}}

\def\of#1{\left(#1\right)}

\def\eq#1{Eq.~(\ref{#1})}
\def\eqs#1#2{Eqs.~(\ref{#1}) and (\ref{#2})}

\def\sof#1{\left[ {#1} \right]}

\def\Ppm{\Psi_\pm}
\def\PApm{\Psi_{A\pm}}
\def\PBpm{\Psi_{B\pm}}
\def\PAp{\Psi_{A+}}
\def\PBp{\Psi_{B+}}
\def\PAm{\Psi_{A-}}
\def\PBm{\Psi_{B-}}
\def\PCp{\Psi_{C+}}

\def\sc{{\text{sc}}}
\def\mod{{\text{mod}}}

\begin{document}

\title{Reconciling Semiclassical and Bohmian Mechanics: \\
II. Scattering states for discontinuous potentials}

\author{Corey Trahan and Bill Poirier}
\affiliation{Department of Chemistry and Biochemistry, and
         Department of Physics, \\
          Texas Tech University, Box 41061,
         Lubbock, Texas 79409-1061}
\email{Bill.Poirier@ttu.edu}

\begin{abstract}

In a previous paper [J. Chem. Phys. {\bf 121} 4501 (2004)]
a unique bipolar decomposition, $\Psi = \Psi_1 + \Psi_2$ was
presented for stationary bound states $\Psi$ of
the one-dimensional \shro\ equation, such that the components
$\Psi_1$ and $\Psi_2$ approach their semiclassical WKB analogs in
the large action limit. Moreover, by applying the Madelung-Bohm
ansatz to the components rather than to $\Psi$ itself, the resultant
bipolar Bohmian mechanical formulation satisfies the correspondence
principle. As a result, the bipolar quantum trajectories are
classical-like and well-behaved, even when $\Psi$ has many nodes, or
is wildly oscillatory. In this paper, the previous decomposition
scheme is modified in order to achieve the same desirable properties
for stationary scattering states. Discontinuous potential systems
are considered (hard wall, step, square barrier/well), for which the
bipolar quantum potential is found to be {\em zero} everywhere,
except at the discontinuities. This approach leads to an exact
numerical method for computing stationary scattering states of any
desired boundary conditions, and reflection and transmission
probabilities. The continuous potential case will be considered in a
future publication.

\end{abstract}

\maketitle                 


\section{INTRODUCTION}

\label{intro}

Much attention has been directed by theoretical/computational
chemists towards developing reliable and accurate means for solving
dynamical quantum mechanics problems---i.e., for obtaining solutions
to the time-dependent \shro\ equation---for molecular systems.
Insofar as ``exact'' quantum methods are concerned, two traditional
approaches have been used: (1) representation of the system
Hamiltonian in a finite, direct-product basis set; (2)
discretization of the wavefunction onto a rectilinear grid of
lattice points over the relevant region of configuration
space. Both approaches, however, suffer from the
drawback that the computational effort scales exponentially with
system dimensionality.\cite{bowman86,bacic89} Recently, a number of
promising new methods have emerged with the potential to alleviate
the exponential scaling problem once and for all. These include
various basis set optimization
methods,\cite{poirier99qcII,poirier00gssI,yu02b,wangx03b}
and build-and-prune methods,\cite{dawes04} such as those based on
wavelet techniques.\cite{poirier03weylI,poirier04weylII,poirier04weylIII}

On the other hand, a completely different approach to the
exponential scaling problem is to use basis sets or grid points,
that themselves evolve over time. The idea is that at any given
point in time, one need sample a much smaller Hilbert subspace, or
configuration space region, than would be required at all
times---thus substantially reducing the size of the calculation. For
basis set calculations, much progress along these lines has been
achieved by the multi-configurational time-dependent Hartree (MCTDH)
method, developed by Meyer, Manthe and co-workers.\cite{meyer90,manthe92}
More recently, time-evolving grid, or ``quantum trajectory''
methods\cite{lopreore99,mayor99,wyatt99,wyatt01b,wyatt01c,wyatt} (QTMs)
have also been developed, and for certain types of systems,
successfully applied at quite high dimensionalities.\cite{wyatt01c,wyatt}

QTMs are based on the hydrodynamical picture of quantum mechanics,
developed over half a century ago by Bohm\cite{bohm52a,bohm52b} and
Takabayasi,\cite{takabayasi54} who built on the earlier work of
Madelung\cite{madelung26} and van Vleck.\cite{vanvleck28} QTMs are
inherently appealing for a number of reasons. First, they offer an
intuitive, classical-like understanding of the underlying dynamics,
which is difficult-to-impossible to extract from more traditional
fixed grid/basis methods. In effect, quantum trajectories are like
ordinary classical trajectories, except that they evolve under a
modified potential $V+Q$, where $Q$ is the wavefunction-dependent
``quantum potential'' correction. Second, QTMs hold the promise of
delivering exact quantum mechanical results without exponential
scaling in computational effort.  Third, they provide a pedagogical
understanding of entirely quantum mechanical effects such as
tunneling\cite{lopreore99,wyatt} and
interference.\cite{poirier04bohmI,zhao03} They have already been
used to solve a variety of different types of problems, including
barrier transmission,\cite{lopreore99} non-adiabatic
dynamics,\cite{wyatt01} and mode relaxation.\cite{bittner02b}
Several intriguing phase space generalizations have also
emerged,\cite{takabayasi54,shalashilin00,burghardt01a,burghardt01b}
of particular relevance for dissipative
systems.\cite{trahan03b,donoso02,bittner02a,hughes04}

Despite this success, QTMs suffer from a significant numerical
drawback, which, to date, precludes a completely robust application
of these methods. Namely: QTMs are numerically unstable in the
vicinity of amplitude nodes.  This ``node problem'' manifests in
several different ways:\cite{wyatt01b,wyatt} (1) infinite forces,
giving rise to kinky, erratic trajectories; (2)
compression/inflation of trajectories near wavefunction local
extrema/nodes, leading to; (3) insufficient sampling for accurate
derivative evaluations. Nodes are usefully divided into two
categories,\cite{poirier04bohmI} depending on whether $Q$ is formally
well-behaved (``type one'' nodes) or singular (``type two'' nodes).
For stationary state solutions to the \shro\ equation, for instance,
all nodes are type one nodes.  In principle, type one nodes are
``gentler'' than type two nodes; however, from a numerical
standpoint, even type one nodes will give rise to the problems
listed above, because the slightest numerical error in the
evaluation of $Q$ is sufficient to cause instability.

In the best case, the node problem simply results in substantially
more trajectories and time steps than the corresponding classical
calculation; in the worst case, the QTM calculation may fail
altogether, beyond a certain point in time. Several numerical
methods, both ``exact'' and approximate, are currently being developed
to deal with this important problem.  The latter category includes
the artificial viscosity\cite{kendrick03,pauler04} and linearized
quantum force methods,\cite{garashchuk04} both of which have
proven to be very stable.
While such approximate methods may not capture the hydrodynamic
fields with complete accuracy in nodal regions, they do allow for
continued evolution and long-time solutions, often unattainable via
use of a traditional QTM. The ``exact'' methods include the adaptive
hybrid methods,\cite{hughes03}
and the complex amplitude method.\cite{garashchuk04b}
In the adaptive hybrid methods,
for which hydrodynamic
trajectories are evolved everywhere except for in nodal regions, where
the time-dependent Schrodinger equation is solved instead to avoid
node problems. Although they have been applied successfully for some
problems, these methods are difficult to implement numerically,
since not only must the hydrodynamic fields be somehow
monitored for forming singularities, but there must also be an
accurate means for interfacing and coupling the two completely
different equations of motion. The complex amplitude method is
cleaner to implement, but is only exact for linear and quadratic
Hamiltonians.

In a recent paper,\cite{poirier04bohmI} hereinafter referred to as
``paper~I,'' one of the authors (Poirier) introduced a new strategy
for dealing with the node problem, based on a bipolar decomposition
of the wavefunction. The idea is to partition the wavefunction into
two (or in principle, more) component functions, i.e. $\Psi = \Psi_1
+ \Psi_2$. One then applies QTM propagation separately to $\Psi_1$
and $\Psi_2$, which can be linearly superposed to generate $\Psi$
itself at any desired later time. In essence, this works because the
\shro\ equation itself is linear, but the equivalent Bohmian
mechanical, or quantum Hamilton's equations of motion (QHEM) are
not.\cite{poirier04bohmI} In principle, therefore, one may improve the
numerical performance of QTM calculations simply by judiciously
dividing up the initial wavepacket into pieces.

Although bipolar decompositions have been around for quite some
time,\cite{floyd94,brown02} their use as a tool for circumventing
the node problem for QTM calculations is quite recent. Two promising
new exact methods that seek to accomplish this are the so-called
``counter-propagating wave'' method (CPWM),\cite{poirier04bohmI}
and the ``covering function'' method (CFM).\cite{babyuk04}
In the CPWM,
the bipolar decomposition is chosen to correspond to the
semiclassical WKB approximation,\cite{poirier04bohmI} for which all of the
hydrodynamic field functions are smooth and classical-like, and the
component wavefunctions are node-free. Interference is achieved
naturally, via the superposition of left- and right-traveling (i.e.
positive- and negative-momentum) waves. For one-dimensional (1D)
stationary bound states, it can be shown that the resultant bipolar
quantum potential $q(x)$ becomes arbitrarily small in the large
action limit, even though the number of nodes becomes arbitrarily
large. (Note: in accord with the convention established in
\Ref{poirier04bohmI}, upper/lower case will be used to denote the
unipolar/bipolar field quantities).  In the CFM, the idea is to
superpose some well-behaved large-amplitude wave, with the actual
ill-behaved (nodal or wildly oscillatory) wave, so as to ``dilute''
the undesirable numerical ramifications of the latter.

This paper is the second in a series designed to explore the CPWM
approach, introduced in paper~I. As discussed there in greater
detail, there are many motivations for this approach, but the
primary one is to reconcile the semiclassical and Bohmian theories,
in a manner that preserves the best features of both, and also
satisfies the correspondence principle. For our purposes, this means
that the Lagrangian manifolds (LMs) for the two theories should become
identical in the large action limit (Sec.~\ref{scattering}).
As described above, a key benefit of the CPWM decomposition is an
elegant treatment of interference, the chief source of nodes and
``quasi-nodes''\cite{wyatt} (i.e. rapid oscillations) in quantum
mechanical systems. An interesting perspective on the role of
interference in semiclassical and Bohmian contexts is to be found in
a recent article by Zhao and Makri.\cite{zhao03}

Whereas paper~I focused on stationary bound states for 1D systems,
the present paper (paper~II) and the next in the series
(paper~III)\cite{poirier05bohmIII} concern themselves with stationary
scattering states. The CPWM decomposition of paper~I is uniquely
specified for any arbitrary 1D state---bound or scattering---and in
the bound case, always satisfies the correspondence principle.
However, the non-$L^2$ nature of the scattering states is such that
the paper~I decomposition generally does {\em not} satisfy the
correspondence principle in this case. Simply put, the quantum
trajectories and LMs exhibit oscillatory behavior in at least one
asymptotic region (thereby manifesting reflection), whereas the
semiclassical LMs do not. This is not a limitation of the CPWM, but
is rather due to the fundamental failure of the basic WKB approximation
to predict any reflection whatsoever for above-barrier energies, as has
been previously well established.\cite{berry72,froman,heading}  In
semiclassical theory, a modification must therefore be made to the
basic WKB approximation, in order to obtain meaningful scattering
quantities. As discussed in Sec.~\ref{scattering} and in paper~III,
our approach will be to apply a similar modification to the exact quantum
decomposition (actually, a {\em reverse} modification) such that the
correspondence principle remains satisfied, and the two theories
thus reconciled, even for scattering systems.

It will be shown the modified CPWM gives rise to bipolar Bohmian LMs
that are {\em identical} to the semiclassical LMs, regardless of
whether or not the action is large. Put another way, this means
that the bipolar quantum potentials $q$ effectively {\em vanish},
so that the resultant quantum trajectory evolution is {\em completely
classical}. Moreover, the resultant component wavefunctions,
$\Psi_1(x)$ and $\Psi_2(x)$, correspond asymptotically to the
familiar ``incident,'' ``transmitted,'' and ``reflected'' waves of
traditional scattering theory. Thus, the modified CPWM implementation
of the bipolar Bohmian approach provides a natural generalization of
these conceptually fundamental entities {\em throughout all of
configuration space}, not just in the asymptotic regions, as is the
case in conventional quantum scattering theory.

The above conclusions will be demonstrated for both discontinuous and
continuous potential systems, in papers~II and~III, respectively.
Discontinuous potentials---e.g. the hard wall, the step potential,
and the square barrier/well---serve as a useful benchmark for the
modified CPWM approach, because the scattering component waves
(e.g. ``incident wave,'' etc.) in this case {\em are} well-defined
throughout all of configuration space, according to a conventional
scattering treatment. Although this is no longer true for continuous
potentials, the foundation laid here in paper~II can be extended to
the continuous (and also time-dependent) case as well, as described
in paper~III. Additional motivation for the development of a
scattering version of the CPWM, vis-a-vis the relevance for
chemical physics applications, is provided in paper~III. Additional
motivation for the consideration of discontinuous potentials is
provided in Sec.~\ref{scattering} of the present paper.


\section{THEORY}

\label{theory}

\subsection{Background}

\label{background}

\subsubsection{Bohmian mechanics}

\label{Bohmian}

According to the Bohmian formulation,\cite{wyatt,holland} the QHEM
are derived via substitution of the 1D (unipolar) wavefunction ansatz,
\eb
\Psi(x,t) = R(x,t) e^{i S(x,t)/\hbar}
\ee
into the time-dependent \shro\ equation. For the 1D Hamiltonian,
\eb
\hat H = -{\hbar^2 \over 2 m} {\partial^2 \over \partial x^2} + V(x),
\label{hameqn}
\ee
this results in the coupled pair of nonlinear partial differential equations,
\ea{ \frac{\partial S(x,t)}{\partial t} & = & -\frac{S'^{2}}{2m} - V(x)
+ \frac{\hbar^{2}}{2m}\frac{R''}{R}, \nonumber \\
\frac{\partial R(x,t)}{\partial t} & = &
-\frac{1}{m}R'\,S'-\frac{1}{2m}R\,S'',}
where $m$ is the mass, $V(x)$ is the system potential, and primes denote
spatial partial differentiation.

The first of the two equations above is the quantum Hamilton-Jacobi
equation (QHJE), whose last term is equal to
$-Q(x,t)$, i.e. comprises the quantum potential correction.
The second equation is a continuity equation.
When combined with the quantum trajectory evolution equations,
i.e.
\ea{
     P & = & m {dx \over dt} = S',  \nonumber \\
         {d P \over dt} & = & -(V'+Q'), }
the continuity equation ensures that the probability [i.e. density,
$R(x,t)^2$, times volume element] carried by individual quantum
trajectories is conserved over the course of their time evolution.

\subsubsection{CPWM decomposition for stationary states}

\label{bipolar}

In paper I, we derived a unique bipolar decomposition,
\eb
\Psi(x) = \Psi_+(x) + \Psi_-(x), \label{bipolardecomp}
\ee
for stationary eigenstates $\Psi(x)$ of 1D Hamiltonians of the
\eq{hameqn} form, such that:
\begin{enumerate}
\item{$\Ppm(x)$ are themselves (non-$L^2$) solutions to the \shro\ equation,
with the same eigenvalue, $E$, as $\Psi(x)$ itself.}
\item{The invariant flux values, $\pm F$, of the two solutions,
$\Ppm(x)$, equal those of the two semiclassical (WKB) solutions.}
\item{The median of the enclosed action, $x_0$, equals that of the
semiclassical solutions.}
\end{enumerate}
There are other important properties of the $\Ppm(x)$,\cite{poirier04bohmI}
as discussed in Sec.~\ref{intro}, and in \Ref{poirier04bohmI}. Nevertheless,
the above three conditions are sufficient to uniquely specify the
decomposition. In the special case of bound (i.e. $L^2$) stationary
states, the real-valuedness of $\Psi(x)$ implies that the
$\Ppm(x)$ are complex conjugates of each other.

\subsection{Scattering systems}

\label{scattering}

It is natural to ask to what extent the above analysis may be
generalized for scattering potentials. Certainly, $\Psi(x)$ itself
is no longer $L^2$, nor even real-valued, and there are generally
two linearly independent solutions of interest for each $E$, instead
of just one. Condition (1) above poses no difficulty for $\Ppm(x)$,
as these component functions are non-$L^2$ and complex-valued, even
in the bound eigenstate case. In principle, condition (2) is not
difficult either; although the flux value depends on the
normalization of $\Psi$ itself, which is not $L^2$, certain
well-established normalization conventions for scattering states
exist, that can be applied equally well to semiclassical and exact
quantum solutions. There is no action median {\em per se} for
scattering states, as the action enclosed within the $\Psi_+(x)$ and
$\Psi_-(x)$ phase space Lagrangian
manifolds\cite{poirier04bohmI,keller60,maslov,littlejohn92} (LMs) is infinite;
however, the scattering analog of condition (3) is related to the asymptotic
boundary conditions, and it is here that one encounters difficulty.
Moreover, an additional concern is raised by the doubly-degenerate
nature of the continuum eigenstates, namely: should each scattering
$\Psi(x)$ have its {\em own} $\Ppm(x)$ decomposition, or should
there be a single $\Ppm(x)$ pair, from which all degenerate
$\Psi(x)$'s may be constructed via arbitrary linear superposition?

To resolve these issues, we will adopt the same general strategy
used in paper I, i.e. we will resort to semiclassical theory as our
guide, wherever possible. We will also exploit certain special
features of the scattering problem not found in generic bound state
systems, such as the asymptotic potential condition $V'(x) \ra 0$ as
$x \ra \pm \infty$ (where primes denote spatial differentiation),
and its usual implications for scattering theory and
applications.\cite{taylor}

The basic WKB solutions are given by
\eb
\Ppm^\sc(x) = r_\sc(x) e^{\pm i s_\sc(x)/\hbar}, \label{scsoln}
\ee
where
\eb
r_\sc(x) = \sqrt{{m F \over s'_\sc(x)}} \qquad\text{and}
\qquad s'_\sc(x) = \sqrt{2 m \sof{E-V(x)}} \label{scrs}
\ee
The corresponding positive and negative momentum functions,
specifying the semiclassical LMs, are given by $p^\sc_\pm(x) = \pm
s'_\sc(x)$. Equations (\ref{scsoln}) and (\ref{scrs}) apply to both
bound and scattering cases; note that for both, $\Ppm^\sc(x)$ are
complex conjugates of each other. The asymptotic potential condition
ensures that these approach exact quantum plane waves
asymptotically, with the usual scattering interpretations, i.e.
$\Psi_+(x)$ in the $x\ra-\infty$ asymptotic region is the incoming
wave from the left (usually taken to be the incident wave),
$\Psi_+(x)$ as $x\ra\infty$ is the outgoing wave from the left (the
usual transmitted wave), etc.

Insofar as determining the corresponding exact quantum solutions
$\Ppm(x)$, the procedure described in paper I is still appropriate
for bound and semi-bound (i.e. on one side only) states, in that the
results satisfy the correspondence principle globally, as desired
(for semi-bound examples, consult the Appendix).  For true
scattering states, however, this procedure fails, in the sense that
if $\Psi_+(x)$ is chosen to match the normalization and flux of
$\Psi^\sc_+(x)$ in the $x\ra\infty$ asymptote, then it will
necessarily approach a nontrivial linear superposition of
$\Psi^\sc_+(x)$ and $\Psi^\sc_-(x)$ in the $x \ra -\infty$
asymptote, and vice-versa. There is therefore an ambiguity as to how
the corresponding quantum $\Ppm(x)$'s should be defined, i.e. which
asymptotic region should be used to effect the correspondence.
More significantly though, {\em either} choice will result in component
functions $\Ppm(x)$ with substantial interference in one of the two
asymptotic regions. This is due to partial reflection of the exact
quantum scattering states, which is not predicted by the basic WKB
approximation. Thus, in the large action limit, the exact quantum
solutions manifest large-magnitude quantum potentials, $q_\pm(x)$,
and rapidly oscillating field functions $q_\pm(x)$, $r_\pm(x)$, and
$p_\pm(x)$---exactly the undesirable behavior that the CPWM was
introduced to avoid---whereas the corresponding basic WKB functions
are smooth, and asymptotically uniform.

The lack of any partial reflection is a well-understood shortcoming of
the WKB approximation\cite{berry72,froman,heading,poirier03capI}---i.e.,
the basic $\Ppm^\sc(x)$ components, though elegantly constructed
from smooth classical functions $r_\sc(x)$ and $s_\sc(x)$, do
not in and of themselves correspond to any actual quantum
scattering solutions $\Psi(x)$. In light of the bipolar
decomposition ideas introduced in paper I, however,
our perspective is the reverse one: for any {\em actual}
quantum $\Psi(x)$, can one determine an \eq{bipolardecomp}
decomposition such that the resultant $\Ppm(x)$ resemble their
well-behaved semiclassical counterparts, and is such a decomposition unique?
Among other properties,\cite{poirier04bohmI} the $\Ppm(x)$ LM's should become
identical to the semiclassical LM's in the large action limit, so as
to satisfy the correspondence principle. Based on the considerations
of the previous paragraph it is clear that the paper I decomposition
does not achieve this goal, when applied to stationary scattering
states.

We defer a full accounting of these issues---in the context of
completely arbitrary continuous potentials $V(x)$---to
paper~III, wherein it will be
demonstrated how to compute exact quantum reflection and
transmission probabilities (and stationary scattering states) using
only classical trajectories, and without the need for explicit
numerical differentiation of the wavefunction. In the present paper,
we lay the foundation for paper III, by focusing attention onto two
key aspects whose development comprises an essential prerequisite.

First, as the paper III approach treats $V(x)$ as a sequence of
steps,\cite{poirier05bohmIII} the present
paper~II will focus exclusively on the
step potential and related discontinuous potential systems, for
which $V(x) = \text{const}$ in between successive steps. Discontinuous
potentials are important for chemical physics, because they model
steep repulsive wells, and are used in statistical theories of liquids.
Moreover, they hold a special significance for QTM methods, for which they
serve as a ``worst-case scenario'' benchmark. Indeed, conventional QTM
techniques {\em always} fail when applied to discontinuous potentials.
To date, The only such calculations that have been
performed\cite{holland} have computed the quantum potential from a
completely separate time-dependent fixed-grid calculation (the
``analytical approach'')\cite{wyatt} rather than directly from
the quantum trajectories themselves. Even if one {\em could} propagate
trajectories for discontinous systems using a traditional QTM, the
trajectories that would be generated would be very kinky and
erratic,\cite{holland} and a great many time trajectories
and time steps would thus be required.

Second, since the new $\Ppm(x)$ do {\em not} satisfy condition (1),
unlike the paper~I CPWM decomposition, the time evolution of these
two component functions is clearly not
that of the time-dependent \shro\ equation. Moreover, since the
$|\Ppm|^2$ are constant over time [because $|\Psi|^2$ itself is
stationary, and \eq{bipolardecomp} is presumed unique], {\em the two
$\Ppm(x,t)$ time evolutions must be coupled together}. It is essential
that the nature of this coupling be completely understood, in order
that the present approach may be generalized to non-stationary state
situations---e.g. to wavepacket scattering, as will be discussed in
future publications.
The ramifications for QTMs
are equally important.  Accordingly, the present paper focuses on
the QTM propagation of the wavefunction and its bipolar
components---with a keen eye towards generality and physical
interpretation---even though the states involved are stationary.
This approach leads to a pedagogically useful reinterpretation of
``incident,'' ``transmitted,'' and ``reflected'' waves---very
reminiscent of ray optics in electromagnetic theory---which is
applicable much more generally than traditional usage might suggest.

\subsection{Basic applications}

\label{timeevol}

The necessary theory will be developed over the course of a
consideration of various model application systems of increasing
complexity.

\subsubsection{free particle system}

\label{free}

Let us first consider the simplest case imaginable, the free
particle system, $V(x)=0$. In this case, the exact solutions
$\Ppm(x) = \Ppm^\sc(x)$ clearly satisfy the conditions of
Sec.~\ref{bipolar}, and the bipolar quantum potentials $q_\pm(x)$
are zero everywhere. Thus, the bipolar decomposition developed for
bound states in paper I can be used directly with this continuum
system, requiring only the slight modification that arbitrary linear
combinations of $\Psi_+^\sc(x)$ and $\Psi_-^\sc(x)$ are to be
allowed, in order to construct arbitrary scattering solutions
$\Psi(x)$. For convenience, the linear combination coefficients will
from here on out be directly incorporated into the amplitude
functions, $r_\pm(x)$, and phase functions, $s_\pm(x)$, so that
\eq{bipolardecomp} is still correct.

If from all solutions $\Psi(x)$ one considers only that which
satisfies the usual scattering boundary conditions (i.e. incident
wave incoming from the left) then the negative momentum wave
$\Psi_-$ vanishes, and $\Psi(x)=\Psi_+(x)$. There is zero
reflection, and $100\%$ transmission. Put another way, the incident
flux, $\lim_{x\ra -\infty} j_+(x)$, is equal to the transmitted
flux, $\lim_{x\ra +\infty} j_+(x)$, where
\ea {
     j_\pm(x) & = & {\hbar \over 2 i m}
     \sof{\Ppm^*(x) {d\Ppm(x) \over dx} -
              {d\Ppm^*(x)\over dx} \Ppm(x)} \nonumber \\
     & & = \sof{p_\pm(x) \over m} r_\pm^2(x), \label{fluxeq}}
[both flux values are equal to $F$, as in \eq{scrs}].

In the quantum trajectory description, flux manifests as
probability-transporting trajectories, which move along the LMs. For
the boundary conditions described above, there are only  positive
momentum trajectories, moving uniformly from left to right with
momentum $p_+(x) = \sqrt{2mE}$. If a $\Psi_-(x)$ contribution were
present, its trajectories would move uniformly in the opposite
direction [$p_-(x)=-\sqrt{2mE}$.] Since the two components $\Ppm(x)$
are in this case uncoupled, the positive and negative momentum
trajectories would have no interaction with each other.

\subsubsection{hard wall system}

\label{hardwall}

We next consider the hard wall system:
\eb
V(x) = {\cases {0 & for $x \le 0$; \cr
                \infty & for $x>0$. \cr }}
\ee
In the $x\le0$ region, the two $\Ppm(x)$ components are exactly the
same as in the free particle case, except that the $\Psi(0)=0$
boundary condition imposes the additional constraints,
\eb
     s_-(0) = s_+(0) + \pi \mod(2 \pi) \qquad ; \qquad
     r_-(0) = r_+(0). \label{hwconst}
\ee This also results in only {\em one} linearly independent
solution instead of two, i.e. $\Psi(x) \propto \sin(k x)$, with $k =
\sqrt{2 m E}/\hbar$. Regarding the LMs and trajectories, in the
$x<0$ region, these are identical to those of Sec.~\ref{free}, e.g.
the $\Psi_+(x)$ LM trajectories move uniformly to the right, towards
the hard wall at $x=0$.

It is natural to ask what happens when the $\Psi_+(x)$ LM
trajectories actually reach $x=0$. There are two reasonable
interpretations. The first is that the trajectories keep moving
uniformly into the $x>0$ region of configuration space. This
approach treats the hard wall system as if it were the free particle
system, but with the $x>0$ region effectively
ignored.\cite{poirier00qcI} This underscores the fact that unlike
$\Psi(x)$ itself, the individual $\Ppm(x)$ components {\em per se}
are unconstrained at the origin---though the \eq{hwconst} constraint
implies a unique correspondence between the two. This interpretation
also makes it clear that for the hard wall system, the paper I
decomposition is essentially identical to the present decomposition,
as is worked out in detail in the Appendix.

In the second interpretation, the effect of the hard wall at $x=0$
is to cause instantaneous elastic reflection of a $\Psi_+(x)$ LM
trajectory momentum, from $p = p_+ = +\sqrt{2 m E}$ to $p = p_- =
-\sqrt{2 m E}$. Afterwards, the reflected trajectory propagates
uniformly backward, along the $\Psi_-(x)$ LM. In this
interpretation, the trajectories never leave the allowed
configuration space, $x\le0$. However, wavepacket reflection is
essentially achieved via trajectory {\em hopping} from one LM to the
other---not unlike that previously considered, e.g., in the context
of non-adiabatic transitions.\cite{tully71}
The trajectory hopping interpretation is adopted in
the present paper, and in paper III, but the first interpretation
will also be reconsidered in later publications.
Note that for discontinuous potentials---and indeed more
generally\cite{poirier05bohmIII}---one can regard trajectory hopping as the
{\em source} of $\Ppm(x)$ interaction coupling.

For the hard wall case, trajectory hopping only manifests at $x=0$,
the sink of all $\Psi_+(x)$ LM trajectories, and the source of all
$\Psi_-(x)$ LM trajectories. If these trajectories are to be
regarded as one and the same via hopping, then a unique field
transformation for $r$, $s$, and all spatial derivatives, must be
specified. Fortunately, the unique correspondence between
$\Psi_+(x)$ and $\Psi_-(x)$ described above, enables one to do just
that. In particular, \eq{hwconst} specifies the correct
transformations for $r$ and $s$, as transported by the quantum
trajectories. All spatial derivatives of arbitrary orders can then
be obtained via spatial differentiation of \eq{scsoln}---although in
the hard wall case, only the $s'$ condition, $p_-(0) = - p_+(0)$ is
relevant, because all higher order derivatives are identically zero.

Since the magnitudes of the $p$ and $r$ fields associated with a
given quantum trajectory are unchanged as a result of the trajectory
hop, \eq{fluxeq} implies that the incident and reflected flux values
are the same (apart from sign), and so the scattering system
exhibits 100\% reflection and zero transmission (along each LM, the
flux is invariant\cite{poirier04bohmI}). These basic facts of the hard wall
system are of course well understood. The point, though, is that we
have now obtained the information in a time-{\em dependent} quantum
trajectory manner, rather than through the usual route of applying
boundary conditions to time-{\em independent} piecewise component
functions. In other words, \eq{hwconst} now refers to individual
{\em quantum trajectories}, rather than to wavefunctions.

This shift of emphasis is very important, and leads to quite a
number of conceptual and computational advantages. For instance, the
standard description of the hard wall stationary states would
decompose these into plane wave components interpreted as
``incident'' and ``reflected'' waves. This language suggests a
process, or change over time---i.e. a state that is initially
incident, at some later time is somehow transformed into a reflected
state. Nothing in the standard description, however, would seem to
render transparent the usage of such terminology, i.e. $\Psi(x)$ is
stationary, and so the reflected and transmitted components are in
fact {\em both} present for all times.  Of course, a localized
superposition of stationary states, i.e. a wavepacket, may well
exhibit such an explicit transformation over the course of the time
evolution, as such a state is decidedly non-stationary. Indeed,
wavepackets are relied upon by the more rigorous formulations of
scattering theory, in order to justify the use of terms such as
``reflected wave,'' even in a stationary context.\cite{taylor} Such
formulations, though certainly legitimate, seem always to require a
clever use of limits, the subtle distinction between unitary and
isometric transformations, and other esoteric mathematical tricks.

On the other hand, the time-dependent bipolar quantum trajectory
hopping picture presented above provides a physicality to such
language that is immediately apparent. Over the course of the time
evolution, although the wavefunction as a whole is stationary, each
individual {\em trajectory} is first incident from the left, then
collides with the hard wall, and is subsequently reflected back
towards the left (i.e. towards $x\ra-\infty$).  The bipolar quantum
trajectories are all classical, as the bipolar quantum potentials,
$q_\pm(x)$, are zero everywhere except at the wall itself.
Interference arises naturally from the superposition of the two
LMs---i.e., from the trajectories that have already progressed to
the point of reflecting, vs. those that have not reflected yet. In
contrast, since $\Psi(x)$ itself exhibits very substantial
interference, and an infinite number of nodes, the traditional
unipolar QTM treatment would be very ill-behaved, i.e. $R(x)$ would
oscillate wildly in the large $k$ limit, and $Q(x)$ would be
numerically unstable near the nodes. Apart from these important
pragmatic drawbacks, the incident/reflected interpretation of the
quantum trajectories would also be lost.

The bipolar quantum trajectory description of the hard wall system
is very reminiscent of  ray optics, as used to describe the
reflection of electromagnetic waves off of a perfectly reflecting
surface.\cite{jackson} Indeed, much can be gained from applying a
ray optics analogy to quantum scattering applications, especially where
discontinuous potentials are concerned. One can construct a simple
gedankenexperiment as follows. Let $x_L<0$ denote some effective
left edge of the system, well to the left of the interaction region.
At some initial time $t=0$, all trajectories on the positive LM
lying to the right of $x_L$ are {\em ignored}, as is the negative LM
altogether. One then evolves the retained trajectories over time,
and monitors the contribution that just these trajectories make to
the total wavefunction. In some respects, it is as if the point
$x_L$ were serving as the initial wavefront for some incoming wave,
that at $t=0$ had not yet reached the hard wall/reflecting surface.
Of course, if the actual wave were in fact truncated in this
fashion, then the discontinuity in the field functions at the
wavefront would result in a very non-trivial propagation over time,
owing to the high-frequency components implicitly present. For
actual waves, the precise nature of the wavefront is known to have a
tremendous impact on the resultant dynamics.\cite{jackson,brillouin14}
We avoid such complicating details by always interpreting the
``actual wave'' to be the full stationary wave itself, i.e. the
truncation is conceptual only.

In the ray optics analogy, the above situation is like a source of
light located at $x_L$, which is suddenly ``turned on'' at $t=0$. It
takes time for the wavefront to propagate to the reflecting surface,
and additional time for the reflected wavefront to make its way back
to $x=x_L$. Prior to the latter point in time, the evolution of the
truncated electromagnetic wave is decidedly {\em non}-stationary;
afterwards however, a stationary wave is achieved, at least within
the region of interest, $x_L\le x \le 0$, as the wavefront has by
this stage propagated beyond this region. The same qualitative
comments apply to the bipolar quantum case, although of course the
evolution equations are different.

A similar prescription may be used to achieve rudimentary
``wavepacket dynamics,'' even in the context of purely stationary
states. Instead of retaining {\em all} initial trajectories that lie
to the left of $x_L$, one retains only those that lie within some
finite interval. The resulting time evolution is analogous to a
light source that is turned on at $t=0$, and then turned off at some
later time (prior to when the wavefront arrives at the reflecting
surface). The initial ``wavepacket'' has uniform density, and moves
with uniform speed towards the hard wall. Interference fringes then
form after the foremost trajectories have been reflected onto the
negative LM. Eventually, all trajectories within the interval are
reflected, at which point interference ceases (the nodes are
``healed''\cite{wyatt}), uniform density is restored, and the
reflected wave travels with uniform speed in the reverse direction,
back towards the starting point $x_L$. Qualitatively, this behavior
is clearly similar to that undergone by actual wavepackets
reflecting off of barrier potentials.

\subsection{More complicated applications}

\label{complicated}

The ideas described above can be easily extended to more complicated
discontinuous potential systems, such as up- and down-step
potentials, and any combination of multiple steps, e.g. square
barriers and square wells. In paper III, they will even be extended
to arbitrary continuous potentials.\cite{poirier05bohmIII} In every case, the
ray optics analogy from electromagnetic theory may also be extended
accordingly. This approach provides a useful perspective on global
reflection and transmission in scattering systems, and in
particular, demonstrates how such quantities may be obtained from a
single, universal expression for local reflection and transmission.

\subsubsection{step potential system---above barrier energies}

\label{stepabove}

We next consider the step potential system:
\eb
V(x) = {\cases {0 & for $x \le 0$; \cr
                V_0 & for $x>0$, \cr }} \label{steppot}
\ee
Classically, this system exhibits 100\% transmission if
the trajectory energy is above the barrier (i.e. $E>V_0$), and
100\% reflection if the trajectory energy is below the barrier
($E<V_0$). Quantum mechanically, all above barrier trajectories
are found to exhibit partial reflection and partial transmission,
although there is a general increase in transmission probability with
increasing energy. The below barrier quantum trajectories exhibit 100\%
reflection, as in the classical case; however, they also manifest
tunneling into the classically forbidden $x>0$ region. Thus even
quantum mechanically, the the above and below barrier cases must be
handled somewhat differently.

To begin with, we consider the above-barrier case.
Note that the LM's are unbounded in either direction, i.e. the classically
allowed region  extends to both asymptotes, $x\ra \pm \infty$.
Incoming trajectories can therefore originate from either asymptote,
thus giving rise to two linearly independent solutions, $\Psi(x)$.
This is in stark contrast to the hard wall system, for which incoming
trajectories could only originate from $x\ra -\infty$, thus resulting
in only one linearly independent solution for $\Psi(x)$.

In the standard time-independent picture, one starts with the four
piecewise solutions,
\eb
     \PApm(x) = e^{\pm i p_A x/\hbar} \quad\text{and}\quad
     \PBpm(x) = e^{\pm i p_B x/\hbar}, \label{steppieces}
\ee
where region $A$ corresponds to $x\le 0$,region $B$ to
$x \ge 0$. The momenta values are classical, i.e.
\eb
     p_A = \sqrt{2 m E}\quad\text{and}\quad
     p_B = \sqrt{2 m (E-V_0)}. \label{steppees}
\ee
Matching $\Psi(x)$ and $\Psi'(x)$ boundary conditions at $x=0$, and
specifying asymptotic boundary conditions for $\Psi(x)$, then enables
a unique determination of the four complex coefficients $A_\pm$ and $B_\pm$
in
\eb
     \Psi(x) = {\cases {A_+ \PAp(x) + A_- \PAm(x) & for $x \le 0$; \cr
                B_+ \PBp(x) + B_- \PBm(x) & for $x\ge0$, \cr }}.
                \label{stepwhole}
\ee

\begin{figure}
\includegraphics[scale=0.65]{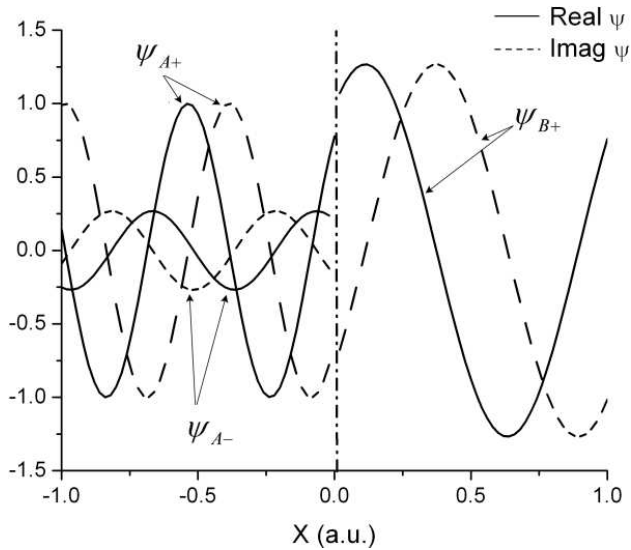}
        \caption{Component waves for a left-incident stationary
eigenstate of the up-step barrier problem with $E>V_0$, as described
in Sec.~\ref{upstep}. Solid and dashed lines represent real and
imaginary contributions, respectively. The dot-dashed line denotes the
location of the step.}
        \label{stepabovefig}
\end{figure}

In general, the solution coefficients depend on the particular
stationary solution of interest. For the usual scattering convention
of an incident wave incoming from the left (Fig.~\ref{stepabovefig})
the solutions are

\ea{
     A_+ = 1 \quad & ; & \quad A_- = R = \of{{p_A - p_B \over p_A + p_B}}
     \nonumber \\
     B_+ = T = \of{{2 p_A \over p_A + p_B}} \quad & ; & \quad B_- = 0,
     \label{stepcoeffs}
} where $R$ and $T$ are respectively, reflection and transmission
amplitudes. When flux is properly accounted for, the resultant
reflection and transmission probababilities (which add up to unity)
are given by \eb
     P_{\text{refl}} = |R|^2 \qquad ;  \qquad
     P_{\text{trans}} = \of{{p_B \over p_A}} |T|^2.
     \label{steprefltrans}
\ee Note that \eqs{stepcoeffs}{steprefltrans} above are correct for
both an ``up-step'' and a ``down-step''---i.e. for $V_0$ positive or
negative. We can also apply these equations to the ``opposite''
boundary conditions, i.e. to an incident wave incoming from the right,
by simply transposing $A$ and $B$ subscripts, and $+$ and $-$ subscripts
($p_A$ and $p_B$ are still positive). This is important, because any
stationary solution $\Psi(x)$ can be obtained as some linear superposition of
left-incident and right-incident solutions.

Regarding the time-dependent interpretation, it is evident that upon
reaching the step discontinuity, left-incident trajectories must be
partially reflected and partially transmitted. The trajectory is
suddenly split into two, one that continues to propagate along the
positive LM for the transmitted $B$ region (i.e. the $B+$ LM) and
the other being instantaneously reflected down to the $A-$ LM.
Moreover, since probability carried by individual quantum
trajectories is conserved,\cite{wyatt,holland} this splitting
must be done in a manner that preserves both probability and flux.
In other words, the local splitting of the trajectory at $x=0$ must
correspond to \eq{stepcoeffs}, which is now regarded as a {\em
local} condition, giving rise to local reflection and transmission
amplitudes, $R$ and $T$. For the present step potential case, these
local quantities are directly related to the global
$P_{\text{refl}}$ and $P_{\text{trans}}$ values via
\eq{steprefltrans}. For multiple step potentials (Sec.~\ref{steps}),
the global expressions above [\eq{steprefltrans}] no longer apply;
however, a local, time-dependent trajectory version of
\eq{stepcoeffs} {\em does} turn out to be correct.

Such an expression, immediately applicable to all single and
multiple step systems, can be written as follows: \ea{
     r_{\text{refl}} = \of{{p_{\text{i/r}} - p_{\text{trans}} \over
              p_{\text{i/r}} + p_{\text{trans}}}}r_{\text{inc}}
               \quad & ; & \quad
     s_{\text{refl}} = s_{\text{inc}}  \label{tdrefl}\\
     r_{\text{trans}} = \of{{2 p_{\text{i/r}} \over
         p_{\text{i/r}} + p_{\text{trans}}}} r_{\text{inc}} \quad & ; & \quad
     s_{\text{trans}} = s_{\text{inc}}. \label{tdtrans}}
In the above equations, ``inc'' refers to any trajectory, locally
incident on some particular step from some particular direction,
which spawns both a locally reflected trajectory, ``refl,'' and a
locally transmitted trajectory, ``trans''. The quantity
$p_{\text{i/r}}$ is the (positive) momentum associated with the
locally incident/reflected trajectory; similarly, $p_{\text{trans}}$
(also positive) is associated with the locally transmitted
trajectory. For above-barrier incident trajectories, note that the
local reflection and transmission amplitudes are both real, thus
ensuring the reality of $r$ and $s$ for the spawned trajectories.

Returning to the step potential system, the ray optics picture can
once again shed some interesting light. The optical analog of the
step is an interface between two media with different indices of
refraction. Light incident on such an interface will partially
reflect back towards the original source, and partially refract
forwards into the new medium. The refraction is completely analogous
to the discontinuous change in momentum, $(p_B-p_A)$, that suddenly
occurs as one crosses the step (Fig.~\ref{trajfig}). In any event,
the ray optics gedankenexperiment described in Sec.~\ref{hardwall}
can also be applied to the step potential system, in order to obtain
a particular stationary solution $\Psi(x)$ with any desired boundary
conditions.

For instance, suppose one is interesting in constructing the
left-incident wave solution, i.e. that of \eq{stepcoeffs}. At $t=0$,
only the $\PAp$ wave is considered, and only those trajectories for
which $x\le x_L$, as before. As the incident trajectories reach the
step, two new waves are dynamically created from the spawned
trajectories: a transmitted wave traveling to the right, and a
reflected wave traveling to the left. A plot of the overall density
$|\Psi(x,t)|^2$ so obtained will change over time, as the
transmitted and reflected wavefronts propagate into their respective
regions (Fig.~\ref{sbbelowfig}). Eventually, however, these
wavefronts will propagate beyond the region of interest, i.e. $x_L
\le x \le x_R$, where $x_R>0$ is the right edge of the region of
interest. When this occurs, the solution for $\Psi(x)$ obtained
within the region of interest will be exactly equal to the
stationary solution with the desired boundary condition.

As in the hard wall case, one can also perform step potential
``wavepacket dynamics'' by restricting consideration to just those
initial $\PAp$ trajectories lying within some coordinate interval.
The wavepacket will propagate towards the step with uniform density
and speed. As the first few trajectories hit the step, a uniform
transmitted wave will be formed in the $B$ region. In the $A$
region, the sudden appearance of a $\PAm$ wave will introduce
interference wiggles in the overall density plot (although no nodes
{\em per se}, owing to partial reflection only). Eventually, after
all trajectories have progressed beyond the step, well-separated
transmitted and reflected wavepackets emerge, propagating in their
respective spaces and directions. There is no longer any
interference in the $A$ region, as the incident wave is now gone,
having been completely divided into the two final contributions.
$P_{\text{refl}}$ and  $P_{\text{trans}}$ values may be determined
via monitors placed at $x_L$ and $x_R$, either by integrating
probability over time as the respective wavepackets travel through,
or by recording the (constant) amplitude values $R$ and $T$, and
applying \eq{steprefltrans}.

\subsubsection{step potential system---below barrier energies}

\label{stepbelow}

The case for which the incident trajectory energies are below $V_0$
requires special discussion. In this case, the classical LMs and
trajectories are confined to the $A$ region only (i.e. to $x\le 0$),
as the entire $B$ region is classically forbidden. In the language
of Sec.~\ref{scattering}, these below barrier states are therefore
semi-bound, implying that there is only one linearly independent
stationary solution, $\Psi(x)$ which without loss of generality,
must be real-valued. This in turn implies that the $\Ppm(x)$ are
complex conjugates of each other, as in the bound state case
discussed in paper~I. Indeed, one option is to simply apply the
paper~I  decomposition to such problems. This approach is discussed
in detail in the Appendix, wherein it is shown to provide a natural
extension of classical trajectories into the tunneling region.

\begin{figure}
\includegraphics[scale=0.75]{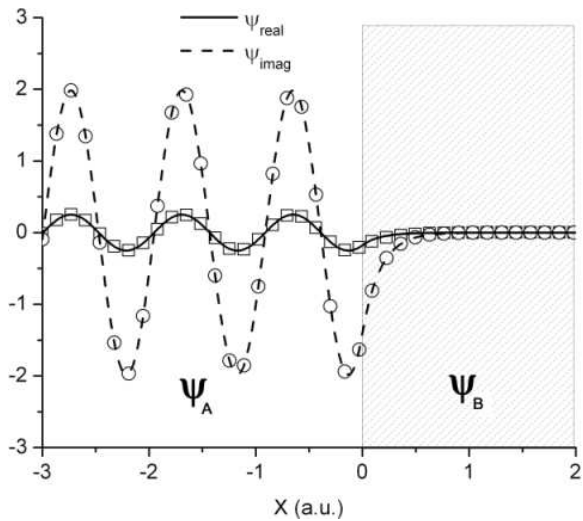}
        \caption{Wavefunction plot for a left-incident stationary eigenstate
of the up-step barrier problem with $E<V_0$, as described
in Sec.~\ref{upstep}. Solid and dashed lines represent real and
imaginary contributions, respectively, for the analytical solution.
Squares and circles denote corresponding numerical results.
The shaded box represents the tunneling region.}
        \label{stepbelowfig}
\end{figure}

On the other hand, the trajectory hopping-based decomposition scheme
offers a different, but also very natural means to accomplish the
same task---which has the added advantage that all bipolar quantum
potentials {\em vanish}, except at $x=0$. The idea is simply to
treat all expressions in Sec.~\ref{stepabove} as being literally
correct for the below barrier case as well, with the understanding
that the requisite quantities need no longer be real. In particular,
$p_{\text{trans}} = i \hbar \kappa$ [\eq{kappaeqn}] becomes pure
positive imaginary, implying that the transmitted trajectories
``turn a corner'' in the complex plane, and start heading off in the
positive imaginary direction, with speed $\hbar \kappa/m$
(Fig.~\ref{complexfig}). Along this path, the transmitted wave is an
ordinary plane wave; however, when analytically continued to the
real axis in the $x>0$ region (via a $90^\circ$ clockwise rotation
in the complex plane), the familiar exponentially damped form
results (Fig.~\ref{stepbelowfig}).

For the reflected wave, \eq{tdrefl} states that the reflected
``phase'' remains unchanged. However, $r_{\text{refl}}$ is now {\em
complex}, leading to an effective phase shift of $2\delta$, where
$\delta$ is defined in the Appendix [\eq{deltaeqn}]. For localized
wavepackets, a physical significance can be attributed to this phase
shift, in both quantum mechanics and electromagnetic theory; it is
the source of the Goos-H\"anchen
effect,\cite{jackson,hirschfelder74} a time delay observed in
conjunction with total internal reflection. Consequently, in the
time-dependent wavepacket context, it may be more appropriate to
associate the phase shift with {\em time}, rather than with $s$ or
$r$---specifically, with the delay time needed to accrue sufficient
action so as to compensate for the shift. For stationary states,
however, such a time delay would be inconsequential, because all
trajectories are identical apart from overall phase. Consequently,
we do not consider such time delays explicitly in this paper, though
we will return to this issue in future publications.

\subsubsection{multiple step systems}

\label{steps}

The most interesting case is that for which there are multiple
discontinuities, occurring at arbitrary locations $x_k$ (with
$k=1,2,\ldots,l$), and dividing up configuration space into $l+1$
regions, labeled $A$, $B$, $C$, etc. In each region, the potential
energy has a different constant value, i.e. $V(x) = V_A$ in region
$A$, etc. From an optics point of view, this system is analogous to
a stack of different materials, each with its own thickness, and
index of refraction. Our primary focus in this paper will be square
barrier/well systems for which $l=2$, and $V_A = V_C$. However, all
of the present analysis extends to the more general case described
above.

In the standard time-independent picture, the solution is obtained
via a straightforward generalization of Eqs. (\ref{steppieces}),
(\ref{steppees}), and (\ref{stepwhole}). However, even when
comparable left-incident boundary conditions are specified as in
Sec.~\ref{stepabove}---i.e. $A_+ = 1$, and (for $l=2$) $C_- =
0$---the remaining coefficient values are fundamentally different
from those of the single-step case. To begin with, only the $l$'th
step exhibits the characteristics of a (locally) left-incident
single-step solution; all other steps involve four non-zero
coefficients, corresponding locally to some superposition of left-
and right-incident waves. Even more importantly, however, the
expressions for the coefficient values as a function of system
parameters {\em in no way} resembles \eq{stepcoeffs}; in particular,
these now depend explicitly on the $x_k$ values, as well as on
$V_A$, $V_B$, etc. The same is also true for the global
$P_{\text{refl}}$ and $P_{\text{trans}}$ expressions, as compared
with \eq{steprefltrans}.

It is this dependence on the other steps that gives rise to the
global nature of the time-independent solutions; i.e. the
coefficient values at one step depend in principle on the properties
of all of the other steps, no matter how far away these might be
located. Consequently, a reflection probability as obtained from the
$A_-$ value associated with the first, $k=1$ step, cannot be
determined without extending the analysis out to the final step at
$x= x_l$, in the standard time-independent picture. On the other
hand, a primary goal of the time-{\em dependent} approach is to
construct a completely {\em local} theory, for which local
reflection and transmission amplitudes associated with any given
trajectory, as it encounters a given step $k$, depend {\em only} on
the properties of the $k$'th step (i.e. on $x_k$, and on the $p$ or
$V$ values to the immediate left and right of $x_k$).  In fact, from
the point of view of the given trajectory, it must be immaterial
whether the potential contains other steps or not---implying that
the correct local relations for the spawned trajectories, if they
exist at all, must be exactly those already specified in
\eqs{tdrefl}{tdtrans}.

\begin{figure}
\includegraphics[scale=0.75]{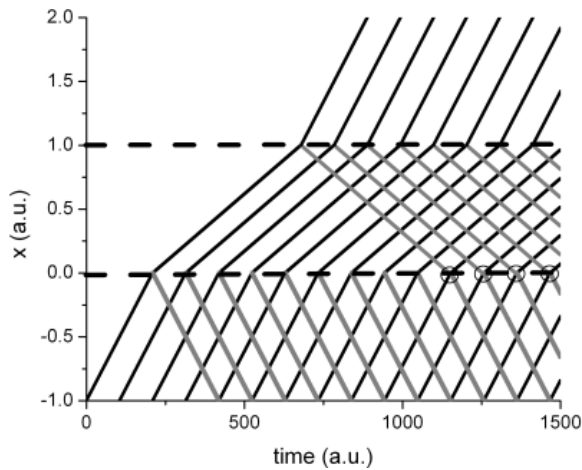}
        \caption{Bipolar trajectory plot for the square barrier problem with
$E>V_0$, as described in Sec.~\ref{squarebarrier}. One trajectory in five is
indicated in the figure. The black/gray solid lines indicate positive/negative
LM trajectories, respectively. The open circles represent recombination
points. The dashed lines denote the two barrier edges, $x_1=0$ and $x_2=1$.}
        \label{trajfig}
\end{figure}

How is it possible that for stationary wavefunctions, whose time
evolution is presumably trivial, an inherently global problem can be
converted to a local one, simply by switching from a
time-independent to a time-dependent perspective? This is because of
the bipolar decomposition, which provides each step with not one,
but two sets of incident trajectories, one from the left, and one
from the right. When there are multiple steps, not only does this
result in a non-trivial superposition for the resultant locally
reflected and transmitted waves, but the trajectories themselves are
subject to multiple spawnings, which effectively enable them to
traverse back and forth over the same regions of configuration space
an arbitrary number of times (Fig.~\ref{trajfig}). This crucial
feature ultimately gives rise to the rich global scattering behavior
observed even in two-step systems. However, it is wholly missed by
any time-independent treatment, even a bipolar one, which can only
summarize the net superposition of all left-traveling and
right-traveling waves.

We now discuss how the local time-dependent theory described above
gives rise to the correct stationary solutions, which is readily
understood by invoking the ray optics description introduced
earlier. For simplicity and definiteness, we consider only the
square potential case, which is optically analogous to say, a single
pane of glass surrounded by vacuum. If a single step gives rise to a
single reflection, then two steps, like a pair of mirrors, results
in an {\em infinite} number of reflections. The same is true of a
pane of glass, within which a single beam of light will be reflected
back and forth at the edges an arbitrary number of times. Of course,
these reflections are not perfect; a portion of the incident flux
always escapes as transmission into the surrounding vacuum.
Consequently, each successive internal reflection is exponentially
damped, in accord with \eq{tdrefl}.

If the globally incident wave is incoming from the left, then at
$x_1$, there are two contributions to $\Psi_{B+}$. One contribution
is the portion of the left-incident $\Psi_{A+}$ wave that is {\em
locally transmitted} through the first step. Apart from a phase
factor, the resultant $B_+$ value would be given by \eq{stepcoeffs}
if this were the only contribution. However, there is also a
contribution that arises from the {\em locally reflected} part of
the right-incident wave, $\Psi_{B-}$. This contribution is zero for
a single step system, but of course non-zero in the multiple step
case.  Although the second contributing wave is right-incident, we
can still use \eqs{tdrefl}{tdtrans} to compute the contribution to
$\Psi_{B+}$, as discussed in Sec.~\ref{stepabove}. For the second,
$k=l=2$ step at $x_2$, there are only left-incident waves;
consequently, $\Psi_{C+}$ and $\Psi_{B-}$ are obtained from a single
source each, i.e. $\Psi_{B+}$ (Fig.~\ref{trajfig}).

\begin{figure}
\includegraphics[scale=0.75]{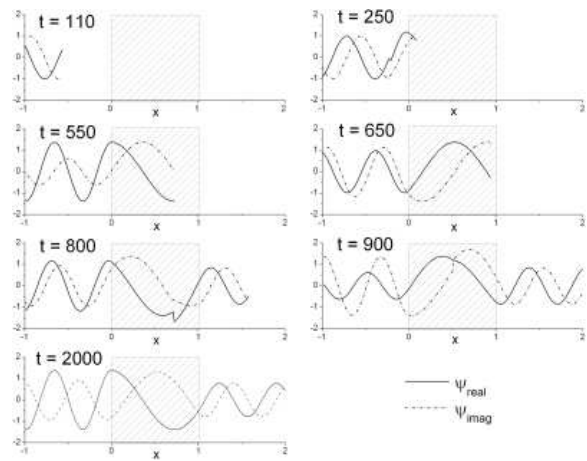}
        \caption{Seven snapshots of the superposition wavefunction,
$\Psi(x,t)$, for the $E>V_0$ square barrier problem, as computed using the
numerical algorithm of Sec.~\ref{numerics}. The shaded box represents the
barrier region. All units are atomic.
The evolving discontinuities found at intermediate times in these
curves denote wavefronts for the newly created reflected/transmitted
components. Over time, the magnitudes of these discontinuities
(i.e. corrections) become arbitrarily small, signifying that numerical
convergence to the left-incident stationary solution has been achieved.}
        \label{sbabovefig}
\end{figure}

The above description refers to the stationary state result,
obtained by our gedankenexperiment in the large time limit only. In
practice this result would be achieved in stages. As in the previous
examples, we imagine that at time $t=0$, one retains only those
trajectories for which $x\le x_L < x_1$. This one-sided trajectory
restriction is somewhat analogous to continuous wave cavity
ring-down spectroscopy.\cite{wheeler98} When the wavefront first hits
the first interface at $x=x_1$, there is partial reflection and
transmission, exactly identical to what would happen for a single
step system. The reflected wavefront propagates beyond the left edge
of the region of interest at $x=x_L$, and for some time, the
reflected amplitude passing through this left edge is constant. The
initially transmitted wavefront eventually reaches the second step
at $x=x_2$ (i.e. the far side of the pane of glass), leading to a
second transmission into the $C$ region, and a second reflection
back through the $B$ region. Eventually, the second transmitted
wavefront reaches the right edge of interest at $x = x_R$, after
which the transmitted amplitude remains constant for some time.

Neither the globally transmitted nor reflected amplitudes for the
times indicated above, as determined via monitors at $x=x_L$ and $x
= x_R$ respectively, are correct. However, we have not yet described
the steady state solution. To do so requires an accounting of the
second reflected wavefront, which eventually reaches the $x=x_1$
step again, this time incident from the right. The resultant locally
transmitted wave becomes an instantaneous second contribution to
$\Psi_{A-}$, and the locally reflected wave plays the same role for
$\Psi_{B+}$. These new contributions give rise to discontinuities in
these waves, that subsequently propagate to the left and right,
respectively (Fig.\ref{sbabovefig}). The new $\Psi_{A-}$ wave
discontinuity eventually reaches $x=x_L$, where it is recorded by
the monitor, giving rise to a sudden change in the reflection
probability value.

The $\Psi_{B+}$ discontinuity propagates to the second step, where
it spawns new discontinuities in $\Psi_{C+}$ and $\Psi_{B-}$. The
former constitutes the border between first- and second-order
transmitted waves, registered at sufficiently later time by the
monitor at $x=x_R$. The latter, second-order $\Psi_{B-}$ wave heads
back towards the first step, to give rise to third-order waves, with
commensurate discontinuities, etc. In principle, this process
continues indefinitely, resulting over time in global transmitted
and reflected waves of arbitrarily high order. However, \eq{tdrefl}
and the relation $P_{\text{refl}}  + P_{\text{trans}} = 1$ ensure
that the result converges to a stationary solution exponentially
quickly. Moreover, since $C_-$ is necessarily zero throughout this
process, it is clear that the stationary state that is converged to
is indeed the one corresponding to the desired boundary condition of
a globally incident wave that is incoming from the left.

Note that in an actual optical system as described above, the
spatial dimensionality is three rather than one, and the incident
wave would usually be taken at some angle to the normal.  If in
addition, the beam has a finite width, then one would observe
separate reflected beams for each order, of exponentially decreasing
brightness. The one-dimensional quantum case, however, is analogous
to a normal incident beam, for which all orders of reflection are
superposed. In addition to providing a pedagogical understanding of
the dynamics that is very much analogous to the optical example
provided, the picture above also suggests a practical numerical
method that may be used to obtain stationary scattering states of
any desired boundary condition (via superposition of globally left-
and right-incident wave solutions, obtained independently).

Note that the ``wavepacket dynamics'' version of the ray optics
analogy may also be applied. In this case, the resultant initial
square wavepacket is somewhat reminiscent of pulsed wave cavity
ring-down spectroscopy.\cite{wheeler98} Once the wavepacket
has penetrated the middle region $B$ (i.e. the pane of glass),
it reflects back and forth between the two edges, with each
reflection giving rise to a left- or right-propagating outgoing
square wavepacket in region $A$ or $C$, and a temporary interference
pattern in region $B$. The amplitude of the central wavepacket
dissipates exponentially in time. All of this complicated
behavior is indeed qualitatively observed in actual wavepacket
dynamics for such systems, but in the present context, is
reconstructed entirely from a single stationary state.


\section{Numerical Details}

\label{numerics}

In this section, we discuss several remaining issues pertaining to
the numerical methods used to generate and propagate the various
bipolar component waves, for the examples discussed in
Secs.~\ref{complicated} and~\ref{results}. For all of these
examples, the numerical algorithm used corresponds to the
gedankenexperiment with one-sided truncation, i.e. to continuous
wave cavity ring-down. In essence, this
consists of just two basic operations:
(1) each piece-wise bipolar component
of the wavefunction [i.e. $\PApm(x)$, $\PBpm(x)$, etc.] is
independently propagated in time over its appropriate region of
space, using the standard QHEMs and QTMs; (2) whenever a
trajectory reaches a turning point, it is immediately deleted, and
replaced with two new trajectories, spawned in the appropriate
locally transmitted and reflected component LMs.

The first operation above, i.e. QTM propagation of the wavefunction
components, is very straightforward. Note that for simplicity, we
have throughout this paper used time-independent expressions for
$\Psi(x)$ and its components, but in reality these evolve over
time---even for stationary states, via $\dot s = \partial s(x,t) /
\partial t = - E$. We have therefore been rather lax in
distinguishing Hamilton's principle function from Hamilton's
characteristic function, although from a trajectory standpoint, it
is always the former that is implied. Since each component is
stationary in its own right, the time evolution of the hydrodynamic
fields is governed by the quantum stationary Hamilton-Jacobi
equation (QSHJE),  rather than the QHJE. Moreover, the fact that the
piecewise $r$ is constant implies that the component quantum
potentials are {\em zero}, resulting in classical HJE's and
trajectories. These conclusions are trivially correct for the
present paper, for which all components are plane waves; however,
the arguments also extend to arbitrary continuous potential
systems.\cite{poirier05bohmIII}

From a numerical perspective, the use of classical trajectories offers many
advantages over a conventional QTM propagation. To begin with, the trajectories
themselves are always smooth if $V(x)$ is smooth, resulting in far fewer
trajectories and larger time steps than would otherwise be the case. Even
more importantly, however, since the quantum potential is not required, there
is no need to compute on-the-fly numerical spatial derivatives of the local
hydrodynamic fields. Consequently, for a given component, the trajectories
are completely independent and need not communicate---again resulting in
fewer of them. Indeed, it is possible to perform an essentially exact
computation using only a {\em single trajectory per wavefunction component}.
This feature is particularly important for the very frontmost trajectory
of the initial ensemble, which for brief periods at later times, will
(via spawning) come to be the {\em only} trajectory to occupy a given
component LM. The subsequent evolution of these lone trajectories does not
require the presence of nearby trajectories.

The spawning of new trajectories, i.e. operation (2) above, also
bears further discussion. In principle, this is always achieved via
application of \eqs{tdrefl}{tdtrans}. For the above barrier case,
$R$ and $T$ are both real, ensuring the reality of $r$ and $s$ for
the spawned trajectories---although in the case of a down step,
$R<0$, resulting in a negative $r_{\text{refl}}$. This is in accord
with the conventions discussed in paper I. However, in this paper,
we find it numerically convenient to adopt the more usual $r>0$
convention. Thus, if \eq{tdrefl} yields a negative
$r_{\text{refl}}$, it is replaced with $-r_{\text{refl}}$, and $\pi$
is added to $s_{\text{refl}}$. A similar, but more complicated
modification is also applied to the below-barrier trajectories, for
which \eqs{tdrefl}{tdtrans} yield complex amplitudes. In this case,
the phase shift is $2 \delta$, as discussed in
Sec.~\ref{stepbelow} and the Appendix.

For a single step system, the algorithm is now essentially complete.
At the initial time $t=0$, a variable number of particles (or
synonymously, grid points) are distributed uniformly along the
$\Psi_{A+}$ manifold, to the left of $x= x_L$. The extent of these
points must be large enough that at the end of the propagation,
there are still $\Psi_{A+}$ grid points that have not yet reached
$x_L$. The grid spacing is mostly arbitrary, but must be small
enough that at sufficiently later times, there is always at least
one trajectory per component LM. The propagation is considered
complete when the reflected and transmitted wavefronts travel beyond
$x_L$ and $x_R$, respectively.

For multiple step systems, the situation is similar, but somewhat
more complex. The primary new feature is the {\em recombination} of
wavefunction components arising from two sources, i.e. from two
locally incident waves coming from opposite directions. The present
algorithm would seem to yield {\em two} subcomponent wavefunctions
for every component, each with its own set of trajectories. If left
``unchecked,'' this would lead to undesirable further
multifurcations for higher orders/later times. A simple solution
would be to propagate each subcomponent long enough that there is at
least one trajectory for each, then extrapolate the corresponding
subcomponent wavefunctions to a common position, where a new
trajectory is constructed for the superposed component wavefunction,
which is then propagated in lieu of the subcomponent trajectories.
This requires dynamical fitting (see below), or at the very least,
extrapolation. Although these numerical operations would be very
stable in the present context, to rule these out altogether as
sources of error in Sec.~\ref{results}, we have adopted a much
simpler approach---i.e. the grid spacing is chosen such that
trajectories from the two component waves incident on a given step
always arrive at the same time (Fig.~\ref{trajfig}). The
corresponding subcomponent wavefunction values are then simply added
together when forming the spawned trajectory. Adopting once again
the $r>0$ convention for the superposed component wave, $\Psi_\pm$,
the corresponding field values are then obtained via
$r=\sqrt{\Psi_\pm^{*}\Psi_\pm}$ and
$s=\arctan\sof{\im(\Psi_\pm)/\re(\Psi_\pm)}$.

As discussed in Sec.~\ref{steps}, multiple step systems allow for
infinite reflections that perpetually modify $|\Psi(x,t)|^2$, in
principle for all time. In practice however, there is exponential
convergence within the region of interest, $x_L \le x \le x_R$, so
that one would not run the calculation indefinitely, but only until
the desired accuracy is reached. Accurate ``error bars'' on the
computed global $P_{\text{refl}}$ and $P_{\text{trans}}$ values are
conveniently provided by the magnitudes of the most recent
discontinuous jumps as recorded by the monitors at $x_L$ and $x_R$.
Note that the number of digits of accuracy scales only {\em
linearly} with propagation time. However, the rate of convergence
depends on the energy value. Near the barrier height, in particular,
convergence may take quite a long time, as the exponent is close to
zero. For all other energies, only a few ``cycles'' should be
required, depending on the level of accuracy desired.

If in addition to reflection and transmission probabilities, the
actual stationary solution over the region of interest is also
desired, then it is necessary to reconstruct $\Psi(x)$. This is
obtained from the final grid, after the propagation is finished,
using a multiple step generalization of \eq{stepwhole}. The first
step is to reconstruct the component wavefunctions $\PApm(x)$,
$\PBpm(x)$, etc., via interpolation or fitting of the hydrodynamic
field values from the corresponding dynamical grid points onto a
much finer common grid (used e.g. for plotting purposes). The second
step is to linearly superpose the $\pm$ components onto the plotting
grid, and to assemble the pieces together over the coordinate range
of interest. For the discontinuous systems considered here, the
number of dynamical grid points per component can be as small as
one---i.e. much smaller, even, than the number of wavelengths! To
our knowledge, such performance has never been achieved previously
by a QTM; however, it does require that the plotting grid be much
finer than the dynamical grid, e.g. at least several points per
wavelength, in order to adequately represent the interference
fringes of the the superposed solution, $\Psi(x)$.


\section{RESULTS}

\label{results}

In this section, we apply the numerical algorithm previously
described to three different applications: the up-step potential,
the square barrier, and the square well.

\subsection{Up-step Potential}

\label{upstep}

The first system considered is the up-step potential, i.e.
\eq{steppot} with $V_0 >0$. Since there are no multiple reflections,
this is in principle a trivial application for the current
algorithm; it therefore serves as a useful numerical test. Both
above barrier (Sec.~\ref{stepabove}) and below barrier
(Sec.~\ref{stepbelow}) energies are considered. We choose
molecular-like values for the constants, i.e. $V_0 = 0.009$ hartree,
and $m=2000$ a.u.
The left and right edges of the region of interest are taken to be
$x_L = -1.0$ a.u. and $x_R = 1.0$ a.u., respectively. At the initial
time, $t=0$, 51 trajectory grid points are distributed uniformly
over the interval $-4 \le x \le -1$ (grid spacing of $0.06$ a.u.).
This number is far greater than what would be needed for dynamical
purposes, but is chosen so as to avoid construction of a separate
plotting grid (Sec.~\ref{numerics}). The hydrodynamic field
functions for the initial $\Psi_{A+}(x)$ wavepacket over the above
interval are taken to be $r(x)=1$ a.u.$^{-1}$ and $s(x) = \sqrt{2mE} x$.

For the above barrier calculation, the energy $E = 2 V_0 = 0.018$
hartree was used. The trajectory propagation and termination were
performed exactly as described in Sec.~\ref{numerics}. The real and
imaginary parts of all three resultant wavefunction components [i.e.
$\PApm(x)$ and $\PBp(x)]$ at the final time, $t=550$ a.u. are
presented in Fig.~\ref{stepabovefig}. All three components exhibit
the desired plane wave behavior, e.g. no interference is evident
within a given component. The resultant $\Psi(x)$ does exhibit
interference in the $A$ region, however, arising from the
superposition of $\PAp$ and $\PAm$.

For the below barrier calculation, the system was given an energy
equal to one half of the barrier height, i.e. $E = V_0/2 = 0.0045$.
As per the discussion in Secs.~\ref{stepbelow} and~\ref{numerics},
tunneling into the forbidden region $B$ is achieved, not through a
quantum potential, but via analytic continuation. At sufficiently
large time ($t=1100$ a.u.) the final wavefunction is reconstructed
from the components, i.e. $\Psi_A(x) = \PAp(x) + \PAm(x)$, and
$\Psi_B(x) = \PBp(x)$. The real and imaginary parts of the
reconstructed wavefunction are presented in Fig.~\ref{stepbelowfig}.
In the figure, squares and circles denote the numerical results
obtained via the present algorithm, whereas the solid and dashed
lines represent the well-known analytic solutions. The agreement is
essentially exact. Note that the real and imaginary parts are in
phase throughout the coordinate range---i.e., $S(x)$ is a constant,
so apart from a phase factor, $\Psi(x)$ is real. Note also that the
tunneling region exhibits the desired exponential decay.

\subsection{Square Barrier}

\label{squarebarrier}

The second system considered is the square barrier. This is a
two-step potential ($l=2$), with $V_A = V_C = 0$, and $V_B = V_0>0$.
The two steps comprise the left and right edges of the barrier, at
$x_1 = 0$ and $x_2=w$, respectively. The constants are chosen as
follows: $V_0 = 0.018$ hartree; $m=2000$ a.u.; $w=1$; $x_L=-1$ a.u.,
$x_R=2$ a.u. Initially, 75 trajectory grid points are distributed
uniformly over the interval $-5 \le x \le  -1$ (grid spacing of
$0.05$ a.u.), which again, is far more than are dynamically
required. The same initial hydrodynamic field functions are used as
in Sec.~\ref{upstep}.

Both above barrier ($E>V_0$) and below barrier ($E<V_0$) energies
are considered. For the above barrier case, $E = 2 V_0 = 0.036$
hartree. Once again, the trajectory propagation and termination were
performed exactly as described in Sec.~\ref{numerics}. In order to
converge $P_{\text{trans}}$ to $10^{-4}$, a propagation time of $3000$
a.u. was required. This corresponds to 3 complete cycles, i.e. a 3rd-order
calculation. Figure~\ref{trajfig}
is a plot of the quantum trajectories for this calculation, in which
every fifth trajectory for each of the five component wavefunctions
is indicated. Trajectory spawning at the two steps is very clearly
evident, as is recombination of pairs of incident waves (indicated
by circles). On the whole, this figure demonstrates all of the
anticipated analogues with ray optics, i.e. parallel trajectories,
reflection and refraction.

In Fig.~\ref{sbabovefig}, the time evolution of the superposition
state $\Psi(x,t)$ is represented, via snapshots of the real and
imaginary parts at seven different times. At $t=0$ a.u., the
incident wavefront is located at $x=-1$ a.u. By $t=250$ a.u., the
wavefront has spawned $\PAm(x)$ and $\PBp(x)$ trajectories; the
former gives rise to the kink (really a discontinuity) somewhat to
the left of the first step. By $t=550$ a.u. and $t=650$ a.u., the
$\PAm(x)$ wavefront has moved outside the region of interest, though
the $\PBp(x)$ wavefront has not quite reached the second step. After
it does so, two new wavefronts are propagated along the $\PCp(x)$
and $\PBm(x)$ LMs (e.g. $t=800$ a.u.), the former of which
propagates beyond the region of interest by $t=900$ a.u. Subsequent
discontinuity magnitudes become exponentially smaller, so that at
sufficiently large time (i.e. $t=2000$ a.u.), the resultant
$\Psi(x)$ has converged to the correct stationary solution.

\begin{figure}
\includegraphics[scale=0.75]{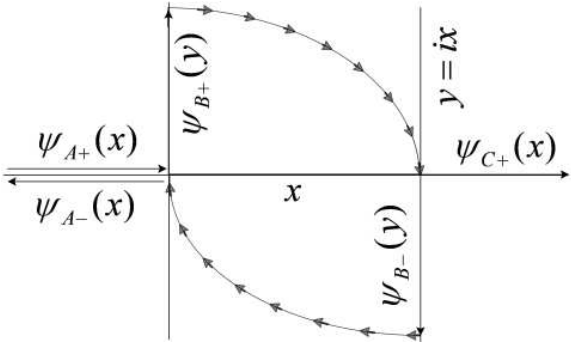}
        \caption{Schematic of the algorithm used to propagate trajectories
into the classically forbidden region of the $E<V_0$ square barrier
problem, as discussed in Secs.~\ref{numerics} and~\ref{squarebarrier}.}
        \label{complexfig}
\end{figure}

For the below barrier case, $E = V_0/2 = 0.009$ hartree, $x_L=-0.5$
a.u., and the other parameters are as above except $w=0.5$ a.u.
Within the barrier, there are in principle an arbitrary number of
reflections back and forth as before. However, substantial amplitude
loss occurs due to tunneling, in addition to partial reflection, as
a result of which fewer cycles are required in order to achieve the
same $10^{-4}$ level of convergence ($t=1400$ a.u., or 2 cycles).
Fig.~\ref{complexfig} indicates how the tunneling dynamics is
achieved. After the wavefront hits
the first step, the transmitted $\PBp$ wave is propagated along the
imaginary axis ($iy$), until the point $y=w$ is reached. When this
occurs, it is necessary to analytically continue $\PBp$ down to the
real axis, in order to compute amplitudes for the new $\PCp(x)$ and
$\PBm(x)$ trajectories that are spawned at the second step. The
latter component propagates in the (negative) imaginary direction,
along $w - iy'$, until $y'=w$, at which point analytic continuation
is once more applied (this time for the first step), and the pattern
repeated.

\begin{figure}
\includegraphics[scale=0.75]{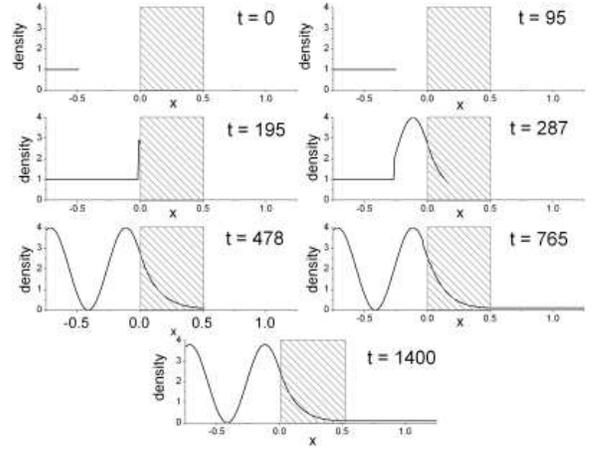}
        \caption{Seven snapshots of the superposition density,
$|\Psi(x,t)|^2$, for the $E<V_0$  square barrier problem,
as computed using the numerical
algorithm of Sec.~\ref{numerics}. The shaded box represents the barrier
region.  All units are atomic. Note that interference manifests in
the incident (left) region only after some incident trajectories
have struck the left barrier edge, causing reflected trajectories
to be created (i.e. just prior to $t=195$, initially). The most advanced
reflected trajectory defines the reflected wavefront, manifesting as
the left-moving discontinuity, e.g. at $t=195$ and $t=287$.}
        \label{sbbelowfig}
\end{figure}

Seven snapshots of the superposition density, $|\Psi(x,t)|^2$, are
displayed in Fig.~\ref{sbbelowfig}. The initial density---equal to
just the $|\PAp(x)|^2$ density---is uniform. After the wavefront
encounters the first step, a reflected $\PAm(x)$ emerges, giving
rise to clearly evident interference in the region $A$. The
$\Psi_B(x)=\PBp(x)$ wave is at this stage perfectly exponentially
damped. Upon encountering the second step, a second contribution,
$\PBm(x)$ emerges; however, this first-order correction is already
extremely small, owing to the large amount of tunneling that has
occurred. The global transmitted wave, $\PCp(x)$, though small, is
clearly seen to have uniform density.

\begin{figure}
\includegraphics[scale=0.75]{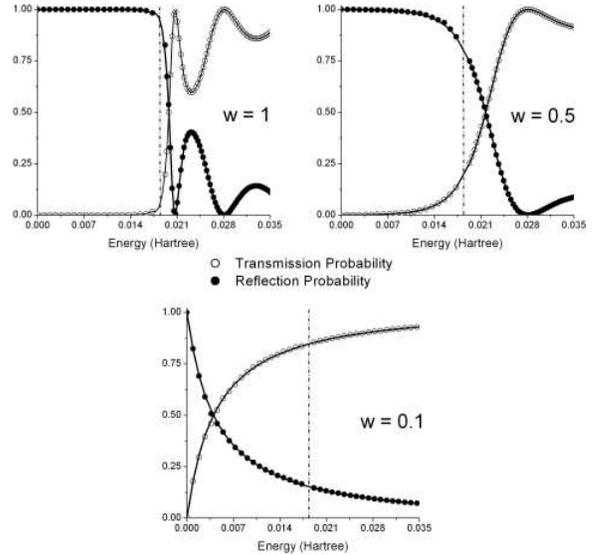}
        \caption{Transmission and reflection probabilities as a function
of energy, for square barrier potentials of three different widths, $w$,
as discussed in Sec.~\ref{squarebarrier}. Solid lines denote analytical
results; open/closed circles denote numerical results, as obtained
via algorithm of Sec.~\ref{numerics}. The vertical dot-dashed lines
represent the barrier height, i.e. $E=V_0$.}
        \label{sbtransreflfig}
\end{figure}

In addition to the two detailed trajectory calculations described
above, we computed $P_{\text{refl}}$ and $P_{\text{trans}}$ for a
large range of $w$ and $E$ values, so as to fully explore (without
loss of generality) the entire range of the square barrier problem.
The numerical results are presented, and compared with known
analytical values,\cite{gasiorowitz} in Fig.~\ref{sbtransreflfig}.
Two aspects of this study bear comment. First, for all $w$ and $E$
values considered, the computed $P_{\text{refl}}$ and
$P_{\text{trans}}$ values agree with the exact values to within an
error comparable to that predicted by the level of numerical
convergence. In particular, the oscillatory energy dependence is
perfectly reproduced. Second, the closer the barrier peak is
approached from either above or below in energy, the longer the time
required to achieve a given level of convergence, as predicted in
Sec.~\ref{numerics}. In particular, for the calculations closest to
the barrier peak, 5-7 cycles were required in order to approximately
maintain a $10^{-4}$ convergence of the transmission probability.
Although a greater number of particles are required in this case,
this poses no great limitation in practice, since one would
presumably never require a calculation precisely at the peak energy.

\subsection{Square Well}

\label{squarewell}

As the final system, we consider the square-well potential, i.e. the
square barrier but with $V_0<0$. In the scattering state context,
there is no tunneling for this system, but in other respects it
resembles the square barrier. From an optics point of view, the
square well corresponds to a central medium with larger index of
refraction than its surroundings, whereas the square barrier
corresponds to a smaller index of refraction, giving rise to the
possibility of total internal reflection (i.e. tunneling). The
parameters are as in Sec.~\ref{squarebarrier}, except that
$V_0=0.009$ hartree, and three different $w$ values are considered:
$w=2$ a.u., $w=4$ a.u., and $w=16$ a.u.

\begin{figure}
\includegraphics[scale=0.75]{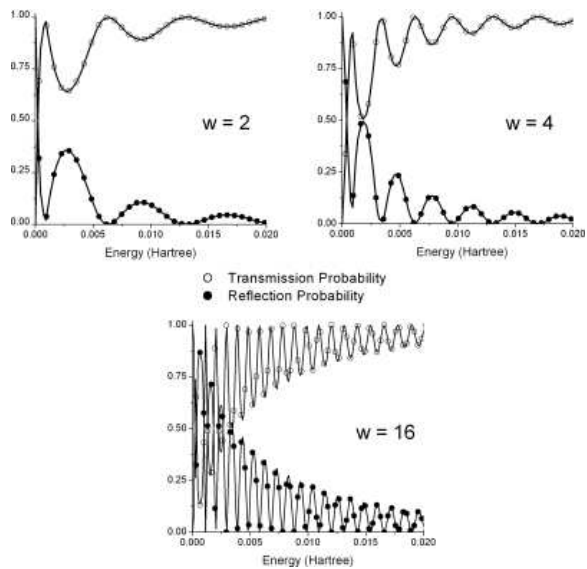}
        \caption{Transmission and reflection probabilities as a function
of energy, for square well potentials of three different widths, $w$,
as discussed in Sec.~\ref{squarewell}. Solid lines denote analytical
results; open/closed circles denote numerical results, as obtained
via algorithm of Sec.~\ref{numerics}.}
        \label{swtransreflfig}
\end{figure}

As the time evolution and trajectory pictures are similar to those
of the previous sections, we focus only on the $P_{\text{refl}}$ and
$P_{\text{trans}}$ calculations, for which once again, a large range
of energies was considered ($0.0005<E<0.2$ hartree). The number of
initial trajectories ranged from 50 to 200, for the highest to the
lowest energies, respectively. The computed transmission/reflection
probabilities were again converged to $10^{-4}$.
The energy-resolved reflection and transmission probabilities are
presented in Fig.~\ref{swtransreflfig}.

As in the square barrier case, excellent agreement is achieved with
the exact analytical results, i.e. on the order of the level of
convergence. This is true despite the fact that the square well
energy curves are decidedly more oscillatory than the square barrier
curves---particularly for wide barriers, for which the $E$
dependence is very sensitive indeed. One particularly important
feature exhibited by the exact curves is the so-called
Ramsauer-Townsend effect,\cite{gasiorowitz} i.e. the phenomenon of 100\%
transmission and zero reflection, even at very low energies.
This occurs when
$\sin(2k_{B}\,w)=0$, and may be regarded as a purely quantum
mechanical resonance phenomenon. Yet it is reproduced perfectly here
in the bipolar decomposition, using {\em classical trajectories}.
Indeed, the bipolar picture provides an interesting physical
explanation, i.e. the right-incident and left-incident waves of the
first step give rise to spawned contributions to $\PAm(x)$ that
exactly cancel each other out via destructive interference.


\section{SUMMARY AND CONCLUSIONS}

\label{conclusion}

As described in paper~I, the \shro\ equation is linear, yet the
equivalent QHEM---obtained via substitution of the Madelung-Bohm
ansatz into the \shro\ equation---are not. This aspect of Bohmian
mechanics suggests that it can be beneficial, both from a
pedagogical and a computational perspective, to apply a suitable
bifurcation (or ``multifurcation'') to the wavefunction prior to
applying the QHEM. Indeed, following the paper~I CPWM bipolar
decomposition for 1D stationary bound states,\cite{poirier04bohmI} the
quantum trajectories become more well-behaved and classical-like,
in precisely the limit in which there are more nodes, and the usual
unipolar calculation breaks down. Moreover, the resultant component
LMs admit a natural physical interpretation in terms of the
corresponding semiclassical LMs.

In the generalization to the stationary scattering states considered
here, a somewhat different bipolar decomposition is found to be
required. The new decomposition is still unique, at least for
discontinuous potentials. However, the resultant components
$\Ppm(x)$ are no longer solutions to the \shro\ equation in their
own right, as a result of which their time evolution is coupled. The
new scheme---though fundamentally different from the old
one---nevertheless bears a correspondence to a modified version of
semiclassical theory appropriate for scattering systems.
Curiously, this semiclassical modification is
{\em not} simply a higher order treatment in $\hbar$;\cite{berry72} if it
were, the corresponding exact quantum modification considered here
would not exist.

In any event, the new decomposition also gives rise to its own
physical interpretation, specific to the scattering context. In
particular, for right-incident boundary conditions, the left and
right asymptotes of $\Psi_+(x)$ respectively represent incident and
transmitted waves, whereas the left asymptote of $\Psi_-(x)$
represents the reflected wave [the right asymptote of $\Psi_-(x)$
approaches zero]. That these conceptually useful asymptotic bipolar
assignments---found even in the most elementary treatments of
scattering---may be extended {\em throughout configuration space}
[even for continuous potentials (paper~III)] represents an important
leap forward, especially for QTMs.

Another pedagogically and numerically useful development from this
approach is the inherently time-dependent ray optics interpretation
that naturally arises, particularly in the discontinuous potential
context. The ray optics approach is anticipated to be a relevant
guiding force in subsequent generalizations of the present
methodology, i.e. to continuous, multidimensional potential systems,
and---as per the discussion in Secs.~\ref{hardwall}
and~\ref{stepabove}---for non-stationary wavepacket dynamics. This
approach also provides a much simpler trajectory-based explanation
of scattering terminology as applied in a stationary context---e.g.
why the ``reflected wave'' is so-called, despite being present from
the earliest times---than those traditionally used.\cite{taylor}

The discontinuous potential applications considered in this
paper---hard wall (Sec.~\ref{hardwall}), step potential
(Secs.~\ref{stepabove}, \ref{stepbelow}, and~\ref{upstep}), and
square barrier/well (Secs.~\ref{steps}, \ref{squarebarrier},
and~\ref{squarewell})---are significant for several reasons. To
begin with, these are the first discontinuous applications of a
genuine QTM calculation that have ever been performed, to the
authors' knowledge. The singular derivatives associated with
discontinuous potential functions would wreak havoc with standard
numerical differentiation routines. Second, the use of a
time-{\em dependent} method for stationary, or time-{\em independent},
applications, is also significant. Ordinarily, the time dependence of
stationary states is regarded as trivial. In the present context,
this is true in a sense for the hard wall and step potential
systems, because the correct answer is ``built in'' the method
itself. For multiple step systems, however, the dynamical truncated
wave approach, i.e. the gedankenexperiment introduced in
Sec.~\ref{hardwall} and further developed in later sections,
yields decidedly nontrivial results.

In particular, the algorithm uses only {\em single step} scattering
coefficients to obtain global scattering quantities for {\em
multiple step} systems. In effect, the time dependent nature of this
approach allows computation of {\em global} properties using a
completely {\em local} method. Not only were exact quantum results
obtained for a full range of system parameters, but the numerical
resources necessary to achieve this---i.e. the number of
trajectories and time steps---were decidedly minimal. Indeed, the
algorithm lives up to the promise made in paper~I, of performing an
accurate quantum calculation with {\em fewer trajectories than
nodes}---a prospect virtually unheard of in a unipolar context.

In future publications, we will naturally attempt to generalize the
methodology described here and in paper~III, for the type of
multidimensional time-dependent wavepacket dynamics relevant to
chemical physics applications. In this context, the scattering
version of the CPWM decomposition developed here is an absolutely
essential first step, as reactive scattering is the underpinning of
all chemical reactions. Additional discussion and motivation will be
provided in paper~III.

\section*{ACKNOWLEDGEMENTS}

This work was supported by awards from The Welch Foundation
(D-1523) and Research Corporation. The authors would like to acknowledge
Robert E. Wyatt and Eric R. Bittner for many stimulating discussions.
David J. Tannor and John C. Tully  are also acknowledged.

%

\section*{Appendix: Bipolar decomposition of semi-bound states}

As discussed in Secs.~\ref{scattering}, \ref{hardwall},
and~\ref{stepbelow}, semi-bound stationary states in 1D are bounded
on one side only, as a result of which they are real-valued and
singly-degenerate, like bound states. Consequently, they are
amenable to the CPWM bipolar decomposition scheme introduced in
paper~I. In this appendix, we apply this decomposition to two
semi-bound systems: the hard wall system, and the below-barrier
up-step system.

\subsection{Hard wall system}

From \Ref{poirier04bohmI}, the most general bipolar decomposition of a hard
wall stationary state---corresponding to Sec.~\ref{bipolar}
condition (1) only---is found to satisfy \eb -\cot\sof{s(x)/\hbar} =
\of{{m F \over \hbar}} \sof{{-\cot(k x) \over k} + B}, \ee where
$s(x) = s_+(x) = -s_-(x)$, and $r_+(x) = r_-(x)$ is obtained from
$s(x)$ via \eq{scrs} (without ``sc'' subscripts). The arbitrary
parameters $F$ and $B$ are the invariant flux and median action
parameters associated with conditions~(2) and (3),
respectively,\cite{poirier04bohmI} although the definition of $F$ has been
changed slightly to account for the scattering normalization
convention, $r_+(x) = 1$. Note that {\em only} the semiclassical
values for these parameters yields a solution that satisfies the
correspondence principle in the large action (i.e. $k$) limit. In
particular, the choice $B=0$ and $F= \hbar k/m$ yields the desired
semiclassical result, $s(x) = \hbar k x$; all other choices exhibit
undesirable oscillatory behavior in $r_\pm(x)$, $s_\pm(x)$, and
$q_\pm(x)$.

\subsection{Up-step system}

For the hard wall system considered above---which is just the
special case of the up-step potential in the limit $V_0 \ra
\infty$---exact agreement is achieved between semiclassical and
quantum LM's in the $x<0$ region. This is the only region of
interest for the hard wall system; however, for finite $V_0$
values---i.e. for general below-barrier up-step stationary
states---there is of course also tunneling into the forbidden
region, which must be accounted for. The paper~I bipolar
decomposition therefore results in LM's that span the {\em entire}
coordinate range $-\infty < x < \infty$. These LMs are given by the
following analytical expression: \eb p(x) = {\cases {\hbar k  & for
$x \le 0$; \cr
                 {2 \hbar \kappa e^{2 \kappa x} \sin{2\delta} \over
                 1 + e^{4 \kappa x} - 2 e^{2 \kappa x}\cos{2\delta}} & for $x>0$. \cr }},
                 \label{ptun}
\ee
where $\delta$ is given by
\eb
\tan \delta = {\kappa \over k} = \sqrt{\of{{V_0 \over E}} - 1},
\label{deltaeqn}
\ee
and
\eb
\kappa = \sqrt{2 m (V_0 - E)}. \label{kappaeqn}
\ee
The $p(x)$ LM function of \eq{ptun} is continuous everywhere,
including at the potential discontinuity at $x=0$. In the $A$
region, it agrees exactly with the semiclassical solution; in the
$B$ region, it decays exponentially to zero.

The paper~I approach thus yields a very natural way to extend
trajectories into the tunneling region. Note that the quantum
potential in this region is not zero; indeed, it exhibits a
discontinuity at $x=0$ that exactly balances that of $V(x)$ itself,
so that the bipolar modified potential is continuous across the
step. Unlike the above-barrier case, the paper~I solution does {\em
not} manifest oscillatory behavior in the large action limit, and so
this approach would at first glance appear to be ideal. There are
two reasons, however, why it is not pursued here. The first reason
is that $r(x)$ diverges asymptotically as $x\ra\infty$, which
according to preliminary numerical investigations, appears to lead
to numerical instabilities for completely QTM-based propagation
schemes. Second, if the barrier were to fall off again at larger $x$
values, so that the tunneling region were finite, then the
asymptotic behavior would be once again undesirably oscillatory.
This would be the case, for example, for the below-barrier energies
of the square barrier system of Sec.~\ref{squarebarrier}.

%
%

%
%
%



%
%


\begin{thebibliography}{10}

\bibitem{bowman86}
J.~M. Bowman, J.~S. Bittman, and L.~B. Harding, J. Chem. Phys. {\bf 85},  911
  (1986).

\bibitem{bacic89}
Z. Ba\v{c}i\'{c} and J.~C. Light, Annu. Rev. Phys. Chem. {\bf 40},  469
  (1989).

\bibitem{poirier99qcII}
B. Poirier and J.~C. Light, J. Chem. Phys. {\bf 111},  4869  (1999).

\bibitem{poirier00gssI}
B. Poirier and J.~C. Light, J. Chem. Phys. {\bf 113},  211  (2000).

\bibitem{yu02b}
H.-G. Yu, J. Chem. Phys. {\bf 117},  8190  (2002).

\bibitem{wangx03b}
X.-G. Wang and T. {Carrington, Jr.}, J. Chem. Phys {\bf 119},  101  (2003).

\bibitem{dawes04}
R. Dawes and T. {Carrington, Jr.}, J. Chem. Phys. {\bf 121},  726  (2004).

\bibitem{poirier03weylI}
B. Poirier, J. Theo. Comput. Chem. {\bf 2},  65  (2003).

\bibitem{poirier04weylII}
B. Poirier and A. Salam, J. Chem. Phys. {\bf 121},  1690  (2004).

\bibitem{poirier04weylIII}
B. Poirier and A. Salam, J. Chem. Phys. {\bf 121},  1704  (2004).

\bibitem{meyer90}
H.-D. Meyer, U. Manthe, and L.~S. Cederbaum, Chem. Phys. Lett. {\bf 165},  73
  (1990).

\bibitem{manthe92}
U. Manthe, H.-D. Meyer, and L.~S. Cederbaum, J. Chem. Phys. {\bf 97},  3199
  (1992).

\bibitem{lopreore99}
C.~L. Lopreore and R.~E. Wyatt, Phys. Rev. Lett. {\bf 82},  5190  (1999).

\bibitem{mayor99}
F.~S. Mayor, A. Askar, and H.~A. Rabitz, J. Chem. Phys. {\bf 111},  2423
  (1999).

\bibitem{wyatt99}
R.~E. Wyatt, Chem. Phys. Lett. {\bf 313},  189  (1999).

\bibitem{wyatt01b}
R.~E. Wyatt and E.~R. Bittner, J. Chem. Phys. {\bf 113},  8898  (2001).

\bibitem{wyatt01c}
R.~E. Wyatt and K. Na, Phys. Rev. E {\bf 65},  016702  (2001).

\bibitem{wyatt}
R.~E. Wyatt, {\em Quantum Dynamics with Trajectories: Introduction to Quantum
  Hydrodynamics} (Springer, New York, 2005).

\bibitem{bohm52a}
D. Bohm, Phys. Rev. {\bf 85},  166  (1952).

\bibitem{bohm52b}
D. Bohm, Phys. Rev. {\bf 85},  180  (1952).

\bibitem{takabayasi54}
T. Takabayasi, Prog. Theor. Phys. {\bf 11},  341  (1954).

\bibitem{madelung26}
E. Madelung, Z. Phys. {\bf 40},  322  (1926).

\bibitem{vanvleck28}
J.~H. {van Vleck}, Proc. Natl. Acad. Sci. U.S.A. {\bf 14},  178  (1928).

\bibitem{poirier04bohmI}
B. Poirier, J. Chem. Phys. {\bf 121},  4501  (2004).

\bibitem{zhao03}
Y. Zhao and N. Makri, J. Chem. Phys. {\bf 119},  60  (2003).

\bibitem{wyatt01}
R.~E. Wyatt, C.~L. Lopreore, and G. Parlant, J. Chem. Phys. {\bf 114},  5113
  (2001).

\bibitem{bittner02b}
E.~R. Bittner, J.~B. Maddox, and I. Burghardt, Int. J. Quantum Chem. {\bf 89},
  313  (2002).

\bibitem{shalashilin00}
D.~V. Shalashilin and M.~S. Child, J. Chem. Phys. {\bf 113},  10028  (2000).

\bibitem{burghardt01a}
I. Burghardt and L.~S. Cederbaum, J. Chem. Phys. {\bf 115},  10303  (2001).

\bibitem{burghardt01b}
I. Burghardt and L.~S. Cederbaum, J. Chem. Phys. {\bf 115},  10312  (2001).

\bibitem{trahan03b}
C.~J. Trahan and R.~E. Wyatt, J. Chem. Phys. {\bf 119},  7017  (2003).

\bibitem{donoso02}
A. Donoso and C.~C. Martens, J. Chem. Phys. {\bf 115},  6309  (2002).

\bibitem{bittner02a}
E.~R. Bittner, J. Chem. Phys. {\bf 115},  6309  (2002).

\bibitem{hughes04}
K.~H. Hughes and R.~E. Wyatt, J. Chem. Phys. {\bf 120},  4089  (2004).

\bibitem{kendrick03}
B.~K. Kendrick, J. Chem. Phys. {\bf 119},  5805  (2003).

\bibitem{pauler04}
D.~K. Pauler and B.~K. Kendrick, J. Chem. Phys. {\bf 120},  603  (2004).

\bibitem{garashchuk04}
S. Garashchuk and V.~A. Rassolov, J. Chem. Phys. {\bf 120},  1181  (2004).

\bibitem{hughes03}
K.~H. Hughes and R.~E. Wyatt, Phys. Chem. Chem. Phys. {\bf 5},  3905  (2003).

\bibitem{garashchuk04b}
S. Garashchuk and V.~A. Rassolov, J. Chem. Phys. {\bf 121},  8711  (2004).

\bibitem{floyd94}
E.~R. Floyd, Physics Essays {\bf 7},  135  (1994).

\bibitem{brown02}
M.~R. Brown, arXiv:quant-ph/0102102  (2002).

\bibitem{babyuk04}
D. Babyuk and R.~E. Wyatt, J. Chem. Phys. {\bf 121},  9230  (2004).

\bibitem{poirier05bohmIII}
B. Poirier, J. Chem. Phys.  , (submitted).

\bibitem{berry72}
M.~V. Berry and K.~V. Mount, Rep. Prog. Phys. {\bf 35},  315  (1972).

\bibitem{froman}
N. Fr{\"o}man and P.~O. Fr{\"o}man, {\em JWKB Approximation} (North-Holland,
  Amsterdam, 1965).

\bibitem{heading}
J. Heading, {\em An Introduction to Phase-integral Methods} (Methuen, London,
  1962).

\bibitem{holland}
P.~R. Holland, {\em The Quantum Theory of Motion} (Cambridge University Press,
  Cambridge, 1993).

\bibitem{keller60}
J.~B. Keller and S.~I. Rubinow, Ann. Phys. {\bf 9},  24  (1960).

\bibitem{maslov}
V.~P. Maslov, {\em Th{\'e}orie des Perturbations et M{\'e}thodes Asymptotiques}
  (Dunod, Paris, 1972).

\bibitem{littlejohn92}
R.~G. Littlejohn, J. Stat. Phys. {\bf 68},  7  (1992).

\bibitem{taylor}
J.~R. Taylor, {\em Scattering Theory} (John Wiley \& Sons, Inc., New York, NY,
  1972).

\bibitem{poirier03capI}
B. Poirier and T. {Carrington, Jr.}, J. Chem. Phys. {\bf 118},  17  (2003).

\bibitem{poirier00qcI}
B. Poirier, Found. Phys. {\bf 30},  1191  (2000).

\bibitem{tully71}
J.~C. Tully, J. Chem. Phys. {\bf 55},  562  (1971).

\bibitem{jackson}
J.~D. Jackson, {\em Classical Electrodynamics}, 2nd  ed. (John Wiley \& Sons,
  New York, NY, 1975).

\bibitem{brillouin14}
L. Brillouin, Ann. Phys. {\bf 44},  177  (1914).

\bibitem{hirschfelder74}
J.~O. Hirschfelder, A.~C. Christoph, and W.~E. Palke, J. Chem. Phys. {\bf 61},
  5435  (1974).

\bibitem{wheeler98}
M.~D. Wheeler, S.~M. Newman, A.~J. Orr-Ewing, and M.~N.~R. Ashfold, JJ. Chem.
  Soc. Faraday Trans. {\bf 94},  337  (1998).

\bibitem{gasiorowitz}
S. Gasiorowitz, {\em Quantum Physics} (John Wiley \& Sons, New York, NY, 1974).

\end{thebibliography}
\end{document}